\documentclass{iopart}
\pdfoutput=1
\usepackage{graphicx}
\usepackage{multirow}
\usepackage{rotating}
\usepackage{subfigure}
\usepackage{iopams}  
\usepackage{bm}
\usepackage{dcolumn}
\usepackage{epsfig}
\usepackage[T1]{fontenc}
\usepackage{graphicx}
\usepackage{graphics}
\usepackage{hyperref}
\usepackage[latin1]{inputenc}
\usepackage{latexsym}
\usepackage{multirow}
\usepackage{rotating}

\newcommand{\rsu}{r^{\ast}}
\newcommand{\drsu}{\dot{r}^{\ast}}
\newcommand{\rsup}{r^{\ast}_{p}}

\newcommand\be{\begin{equation}}
\newcommand\ba{\begin{eqnarray}}
\newcommand\ee{\end{equation}}
\newcommand\ea{\end{eqnarray}}

\newcommand{\mb}[1]{\mbox{\boldmath $#1$}}
\newcommand{\met}{\mbox{g}}

\newcommand{\HH}{{\mbox{\tiny H}}}
\newcommand{\II}{{\mbox{\tiny I}}}
\newcommand{\LL}{{\mbox{\tiny L}}}
\newcommand{\PP}{{\mbox{\tiny P}}}
\newcommand{\RR}{{\mbox{\tiny R}}}
\newcommand{\SF}{{\mbox{\tiny SF}}}
\newcommand{\SI}{{\mbox{\tiny S}}}
\newcommand{\TT}{{\mbox{\tiny T	}}}
\newcommand{\MAX}{{\mbox{\tiny max}}}

\begin{document}

\title[Tuning Time-Domain Pseudospectral Computations of the Self-Force]
{Tuning Time-Domain Pseudospectral Computations of the Self-Force on a Charged Scalar Particle}

\author{Priscilla Canizares and Carlos F. Sopuerta}

\address{Institut de Ci\`encies de l'Espai (CSIC-IEEC), Facultat de Ci\`encies, Campus UAB, 
Torre C5 parells, Bellaterra, 08193 Barcelona, Spain}

\ead{pcm@ieec.uab.es, sopuerta@ieec.uab.es}

\begin{abstract}
The computation of the self-force constitutes one of the main challenges for the construction 
of precise theoretical waveform templates in order to detect and analyze extreme-mass-ratio 
inspirals with the future space-based gravitational-wave observatory LISA.  Since the number 
of templates required is quite high, it is important to develop fast algorithms both for the 
computation of the self-force and the production of waveforms.  In this article we show how to 
tune a recent time-domain technique for the computation of the self-force, what we call the 
Particle without Particle scheme, in order to make it very precise and at the same time very 
efficient.  We also extend this technique in order to allow for highly eccentric orbits.   
\end{abstract}

\pacs{04.20.-q,04.25.Nx,04.30.Db,02.70.Hm}
\ams{35Q75,65M70,83C10}
\submitto{\CQG}

\maketitle

\section{Introduction}  \label{intro}
Gravitational Wave Astronomy has the potential to unveil the secrets of physical phenomena
not yet understood involving strong gravitational fields.  In order to realize this
potential there are observatories/detectors either operating or being design/constructed
that cover a significant part of the relevant gravitational-wave frequency spectrum.
From the very low band ($10^{-9}-10^{-6}$ Hz), where we have the pulsar timing arrays to 
the high frequency band ($1-10^{4}$ Hz), where most ground-based detectors operate. In low 
frequency band ($10^{-4}-1$ Hz), not accessible from the ground, we find the future Laser 
Interferometer Space Antenna (LISA)~\cite{LISA}, an ESA-NASA mission consisting of three 
spacecrafts forming a triangular constelations in heliocentric orbit 
(see~\cite{Danzmann:2003ad,Prince:2003aa} for details).  The targets of LISA are: (i) massive 
black hole (BH) mergers; (ii) the capture and subsequent inspiral of stellar compact objects 
(SCO) into a massive BH sitting at a galactic center; (iii) galactic compact binaries; and 
(iv) stochastic backgrouns of diverse cosmological origin.  

For many of these systems it is crucial to have a priori gravitational waveform models with
enough precision to extract the corresponding signals from the future LISA data stream, and
also to estimate the physical parameters of the system with precision.  Here we focus in the
second type of LISA sources, namely the so-called extreme-mass-ratio inspirals (EMRIs).  The
EMRIs of interest for LISA consists of massive BHs with masses in the range $M= 10^4-10^7 M_{\odot}$,
and SCOs with masses in the range $m = 1-50 M_{\odot}$.  The SCO moves around the massive BH following
highly relativistic motion well inside the strongest field region of the BH.  The orbit is not
exactly a geodesic of the BH because of the SCO own gravity, which influences its own motion,
making the orbit to shrink until it plunges into the BH.   During the last year before plunge,
it has been estimated~\cite{Finn:2000sy} that the SCO spends of the order of $10^{5}$ cycles 
(depending on the type of orbit) inside the LISA band.   As a consequence the EMRI GWs carry a 
detailed map of the BH geometry.  This will allow, in particular, to test the spacetime geometry 
of BHs and even alternative theories of gravity 
(see, e.g.~\cite{Schutz:2009zz,Sopuerta:2010zy,Babak:2010ej}).  
We can also understand better the stellar dynamics near galactic nuclei, populations of 
stellar BHs, etc. (for a review see~\cite{AmaroSeoane:2007aw}).

All this requires very precise EMRI gravitational waveforms (the precision of the phase should 
be of the order of one radian per year).  Due to the extreme mass ratio of these systems, 
$\mu=m/M \sim 10^{-7} -10^{-3}$, we can describe them in the framework of BH perturbation 
theory, where the SCO is modeled as a point-like mass and the backreaction effects are described
as the effect of a local force, the so-called {\em self-force}.  This self-force is essentially
determined by the derivatives of the metric perturbations (retarded field) at the particle location, 
which need to be regularized due to the singularities introduced by the particle description.  
The calculation of the retarded field has to be done numerically, either in the frequency or time
domain.  In this work we concentrate on a recent approach to self-force calculations in the time
domain introduced in~\cite{Canizares:2008dp,Canizares:2009ay} for circular orbits and extended to 
eccentric orbits in~\cite{Canizares:2010yx}.  This is a multidomain approach
in which the point particle is always located at a node between two domains, and hence it has been
named the Particle without Particle (PwP) scheme.  The main advantage is that the equations to be
solved are homogeneous and the particle location enters via junction/matching conditions.  In this
way we are dealing with smooth solutions at each domain, which is crucial for the convergence of
the numerical method used to implement the method, the PseudoSpectral Collocation (PSC) 
method in our case.

In this article we describe how to tune the numerical implementation of the PwP scheme
to achieve high precision results with modest computational resources.  We focus on two fronts: 
(i) How to change the framework introduced in~\cite{Canizares:2009ay,Canizares:2010yx} to compute the
self-force for orbits with high eccentricity, and (ii) how to pick the size of the computational 
domain in order to make our computations much more efficient, so that we can achieve very precise 
results with a modest amount of computational resources.  A similar analysis for a different 
computational technique has been presented in~\cite{Thornburg:2010tq}.

The plan of the paper is the following: In Sec.~\ref{scalarparticlearoundsch} we describe briefly 
the foundations of self-force calculations in a simplified model based on a charged particle 
orbiting a non-rotating BH.  In Sec.~\ref{pwpscheme} we summarize the PwP scheme and extend the 
multidomain structure in order to allow for computations in the case of high eccentric orbits.  
We also show the convergence properties of the PSC method numerical implementation.  
In Sec.~\ref{fitmod} we describe how to choose the resolution 
depending on the mode in order to perform optimal calculations in the sense of computational resources.  
We finish with conclusions and a discussion in Sec.~\ref{discussion}, where we also present some 
numerical results.

\section{Scalar charged particle orbiting a non-rotating BH}\label{scalarparticlearoundsch}

The dynamics of a scalar charged particle is governed by the equation of motion of the 
associated scalar field (see, e.g.~\cite{Poisson:2004lr}):
\begin{equation}
\met^{\alpha\beta}\nabla^{}_{\alpha}\nabla^{}_{\beta}\Phi(x)= -4\pi q\int^{}_{\gamma}  
d \tau\, \delta^{}_4(x,\textit{z}(\tau))  \,, \label{retfield}
\end{equation}
and its own equation of motion
\begin{equation}
m \,\frac{du^{\mu}}{d\tau} = F^{\mu} = q \,(\met^{\mu\nu} + u^{\mu}u^{\nu}) \left. 
\left(\nabla^{}_{\nu}\Phi\right) 
\right|^{}_{\gamma} \,, ~~ u^{\mu} = \frac{dz^{\mu}}{d\tau} \,. \label{particlemotion}
\end{equation}
where $m$, $q$, and $\gamma$ are the particle mass, charge, and timelike worldline 
(parameterized as $x^{\mu}=z^{\mu}(\tau)$, being $\tau$ proper time) respectively,
$\met^{\alpha\beta}$ is the spacetime inverse metric (the BH metric, the Schwarzschild
metric in our case), $\nabla^{}_{\mu}$ the associated 
canonical connection, and $\delta^{}_{4}(x,x')$ is the invariant Dirac delta distribution.
These two equations are coupled [Eq.~(\ref{retfield}) is a partial differential equation whereas
Eq.~(\ref{particlemotion}) is an ordinary differential equation].  Eq.~(\ref{retfield}) 
describes the retarded scalar field generated by the particle and Eq.~(\ref{particlemotion})
how this field affects in turn the particle trajectory, mimicking the backreaction mechanism
that produces the inspiral of EMRIs in the gravitational case (see~\cite{Barack:2009ux,Poisson:2004lr} for
reviews on the self-force problem).

Since we are dealing with a non-rotating BH, using the spherical symmetry we can 
expand the retarded field in scalar spherical harmonics
\begin{eqnarray}
\Phi(x^\mu) = \sum_{\ell=0}^{\infty}\sum_{m=-\ell}^{\ell}\Phi^{\ell m}(t,r)
Y^{\ell m}(\theta,\varphi)\,, \label{phiex}
\end{eqnarray}
so that each harmonic mode $\Phi^{\ell m}(t,r)$ satisfies a $1+1$ wave-type equation:
\begin{eqnarray}
\left\lbrace  -\frac{\partial^2}{\partial t^2} + \frac{\partial^2}{\partial \rsu{}^{2}} 
-V^{}_{\ell}(r) \right\rbrace(r\,\Phi^{\ell m})= S^{\ell m}\delta (r-r^{}_{p}(t))\,,  \label{master}
\end{eqnarray}
where $r^{}_{p}(t)$ is the radial trajectory of the particle, $\rsu = r + 2M \ln(r/2M-1)$  
is the radial \emph{tortoise} coordinate, $M$ is the BH mass, and $V^{}_{\ell}(r)$ is the 
Regge-Wheeler potential for scalar fields
\be
V^{}_{\ell}(r) = f(r) \left[ \frac{\ell(\ell+1)}{r^2}+\frac{2M}{r^3}\right]\,,\qquad
f(r)=1-\frac{2M}{r}\,,
\ee
where $S^{\ell m}$ is the coefficient of the distributional source term due to the
presence of the particle given by
\begin{eqnarray}
S^{\ell m} = -\frac{4\pi qf(r^{}_{p})}{r^{}_{p}u^t}\,
\bar{Y}^{\ell m}\left(\frac{\pi}{2},\varphi^{}_{p}(t)\right)\,,
\label{source}
\end{eqnarray}
where $\varphi^{}_{p}(t)$ is the azimuthal trajectory of the particle and we have assumed, 
without loss of generality, that the orbit takes place at the
equatorial plane $\theta=\pi/2$.  The singular character of this source term makes the 
retarded field to diverge at the particle location. Therefore, in order to obtain
a meaningful self-force through Eq.~(\ref{particlemotion}) we must use a regularization
scheme.  We employ the {\em mode sum} regularization 
scheme~\cite{Barack:1999wf,Barack:2000eh,Barack:2001bw,Barack:2001gx,Mino:2001mq,Barack:2002mha}
(see also~\cite{Detweiler:2002gi,Haas:2006ne}), 
which provides an analytical expression for the singular part of the retarded field mode by mode, 
$\Phi^{\ell m}_{\SI}$, at the particle location.  Hence, by subtracting this singular contribution 
from the full retarded field modes, $\Phi^{\ell m}$ (which are finite at the particle location), 
we obtain a smooth and differentiable field at the particle location, 
$\Phi^{\ell m}_{\RR} = \Phi^{\ell m}-\Phi^{\ell m}_{\SI}$.
Adding all the harmonic modes we obtain the full regular field, $\Phi^{}_{\RR}$,
from which we can obtain the scalar self-force as in Eq.~(\ref{particlemotion}), that is
\begin{equation}
\textit{F}^{\;\mu}_{\SF} = q \,(g^{\mu\alpha} + u^{\mu} u^{\alpha} )\left.\left(
\nabla_{\alpha}\Phi^{}_{\RR}\right)\right|^{}_{\gamma}\,. 
\label{self_force}
\end{equation}
%

\section{Formulation of the PwP scheme}\label{pwpscheme}

From the previous discussion it is clear that the main challenge is the computation
of the harmonic modes of the retarded scalar field.   This has to be done numerically and
here we describe the scheme and numerical implementation that we use.   We solve 
Eq.~(\ref{master}) in the time domain.  To that end we use a multi-domain framework in
which the particle is located always at a node in the interface between two domains,
what we call the PwP scheme.  The main advantage of the PwP scheme is that it avoids the
presence of the singularity associated with the particle in the scalar field equations,
which is crucial for the convergence properties of the numerical method.  The particle 
then enters as junction conditions of the different domains.
This idea was already used in~\cite{Sopuerta:2005gz} for
the computation of energy and angular momentum fluxes for all kinds of orbits.  The
scheme was used for the computation of the self-force for the first time in~\cite{Canizares:2009ay}
for the circular case, and extended in~\cite{Canizares:2010yx} to the eccentric case.  
Nevertheless, the scheme as presented in~\cite{Canizares:2010yx} can be used for moderate
eccentricities, typically $e \lesssim 0.5$.  The reason for this is that in the eccentric
case the one-dimensional spatial grid needs to be changed in time in order to keep the
particle at a boundary node.  In~\cite{Canizares:2010yx} only the domains adjacent to the
particle change in time, which limits the eccentricity of the orbits.  This limitation
is related to the turning points of the orbital motion (pericenter and apocenter), where
one of the domains is effectively much larger than the other.  In the large domain the 
resolution will be poor for high eccentricities, whereas in the small domain it will be
high but this imposes a very restrictive Courant-Friedrichs-Lax (CFL) condition on the time
step, specially in the case of PSC methods, where the CFL condition is proportional to 
$1/N^{2}$, being $N$ the number of collocation points per domain.  Therefore, pushing the
scheme as it is for high eccentricities means to increase dramatically the computational
resources required to obtain precise estimations of the self-force.

\begin{figure*}
\centering
\includegraphics[width=0.85\textwidth]{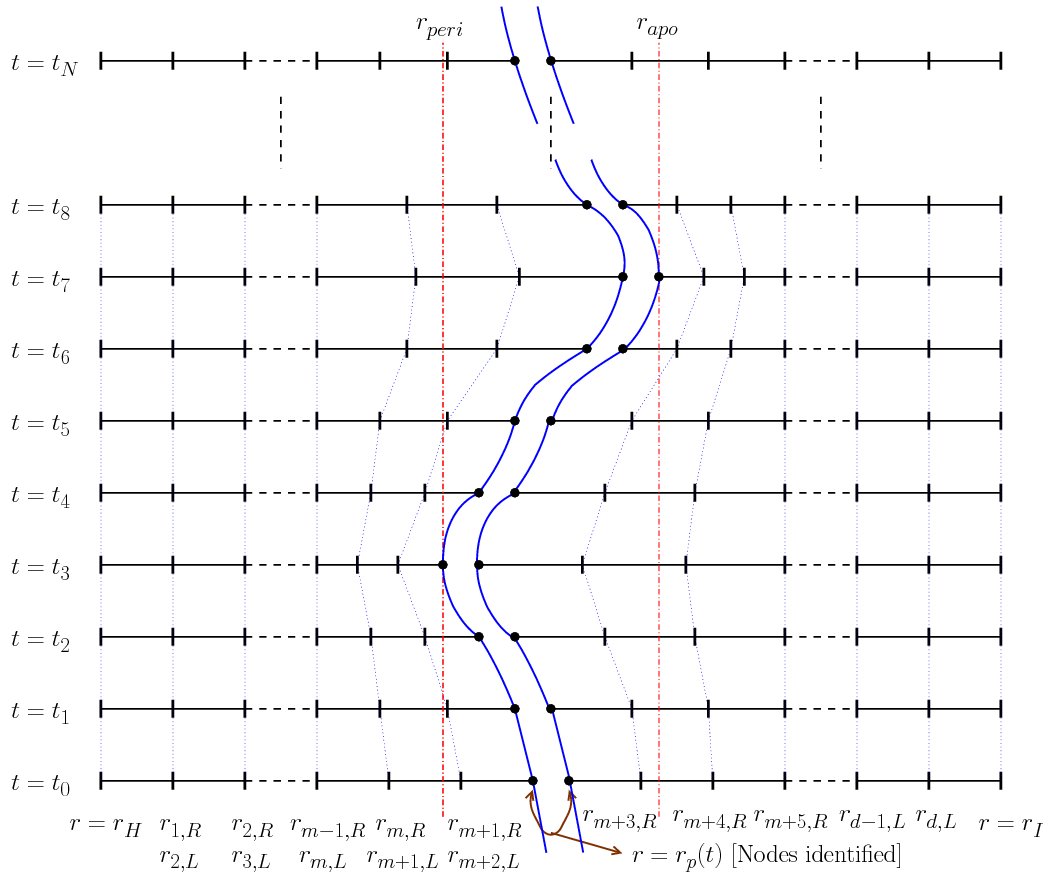}
\caption{Structure of the one-dimensional spatial computational domain 
for a generic orbit.   The trajectory is bounded to the interval between 
the pericenter ($r^{}_{peri}$) and the apocenter ($r^{}_{apo}$). 
In the setup shown there are three dynamical domains to the left and right of the
particle, namely $[\rsu_{m,\LL},\rsu_{m,\RR}]\cup[\rsu_{m+1,\LL},\rsu_{m+1,\RR}]\cup[\rsu_{m+2,\LL},\rsu_{m+2,\RR}]$
(left)
and $[\rsu_{m+3,\LL},\rsu_{m+3,\RR}]\cup[\rsu_{m+4,\LL},\rsu_{m+4,\RR}]\cup[\rsu_{m+5,\LL},\rsu_{m+5,\RR}]$ (right)
with $\rsu_{m+2,\RR}=\rsu_{m+3,\LL}=\rsup$.   The figure illustrates how the dynamical domains change
their coordinate size to adapt to the particle motion.} 
\label{multidomain}
\end{figure*}

Here we reformulate the PwP to allow for such high eccentricity orbits, which are 
of interest for the EMRIs that LISA is expected to observe (see, e.g.~\cite{AmaroSeoane:2007aw}).
The one-dimensional spatial domain $\rsu\in(-\infty,+\infty)$ is truncated from both sides, so 
that the global computation domain is $\Omega=\left[\rsu_{\HH},\rsu_{\II}\right]$, where 
$\rsu_{\HH}$ corresponds to the truncation in the direction to the BH horizon and $\rsu_{\II}$ 
corresponds to the truncation in the direction to spatial infinity.  Now, we divide $\Omega$
in a set of subdomains: $\Omega = \cup_{a=1}^{d}\Omega^{}_{a}$, where 
$\Omega^{}_{a} = \left[ \rsu_{a,\LL}, \rsu_{a,\RR}\right]$ with $\rsu_{a,\RR} = \rsu_{a+1,\LL}$
($a=1,\ldots,d-1$).  We consider two types of domains,
{\em static} ($\drsu_{a,\LL}=\drsu_{a,\RR}=0$) and {\em dynamic} (either $\drsu_{a,\LL}\neq 0$ or 
$\drsu_{a,\RR}\neq0$).  For instance, the domains surrounding the particle, 
$\Omega^{}_{\PP-1} = \left[ \rsu_{\mbox{\tiny {\rm P}-1},\LL}, \rsu_{p}\right]$ and 
$\Omega^{}_{\PP} = \left[ \rsu_{p}, \rsu_{\PP,\RR}\right]$ 
(where P is the number of the domain with the particle at the left node, $\Omega^{}_{\PP}$,
and hence $\rsu_{\mbox{\tiny {\rm P}-1},\RR}=\rsu_{\PP,\LL}=\rsu_{p}$), are dynamical.  Then, generalizing the
setup presented in~\cite{Canizares:2010yx}, here we consider an arbitrary number of dynamical
domains.   Those domains are all around the particle as shown in Figure~\ref{multidomain}.
In the case of circular orbits ($r^{}_{p}(t) = const.$) all domains are static.

We discretize the spatial domain using the PSC method (see, e.g.~\cite{Boyd}).  
For this purpose, each physical domain $\Omega^{}_{a}$ ($a=1,\dots,d$) is associated with a {\em spectral}
domain, $[-1,1]$ as we use Chebyshev polynomials as the basis functions
for the spectral expansions, which is discretized independently.  Our choice of discretization is
the {\em Lobatto-Chebyshev} grid, with collocation points given by 
$X^{}_{i} = - \cos(\pi\,i\,/\,N)$ ($i=0,\ldots,N$).  Of course, the physical and spatial
domains have to be mapped and this is the crucial point for our numerical implementation
of the PwP scheme.  For a given domain $\Omega_{a} =\left[\rsu_{a,\LL}, \rsu_{a,\RR}\right]$ 
($a=1,\ldots,d$), the spacetime mapping to the spectral domain is taken to be 
linear
\begin{eqnarray}
\begin{array}{ccccc}
X^{}_{a} & : & \mathbb{T}\times\left[ \rsu_{a,\LL}, \rsu_{a,\RR}\right] & \longrightarrow & \mathbb{T}\times\left[-1,1\right] \\[2mm]
         &   & (t,\rsu)                                 & \longmapsto     &  (T, X^{}_{a}) 
\end{array} 
\end{eqnarray} 
where $\mathbb{T} = [t^{}_{ini},t^{}_{end}]$ is the evolution time interval and
\begin{eqnarray}
T(t)= t\;,\qquad
X^{}_{a}(t,\rsu) = \frac{2\rsu- \rsu_{a,\LL}- \rsu_{a,\RR}}{ \rsu_{a,\RR}-\rsu_{a,\LL} } \,.    
\label{map1}
\end{eqnarray} 
The inverse (linear) mappings from the spectral domain to each of the subdomains $\Omega_{a}$ are given by
\begin{eqnarray}
\begin{array}{ccccc}
\left.\rsu\right|^{}_{\Omega^{}_{a}} & : & \mathbb{T}\times\left[-1,1\right] & \longrightarrow & \mathbb{T}\times
\left[\rsu_{a,\LL},\rsu_{a,\RR}\right] \\[2mm]
                                     &   & (T, X)                            & \longmapsto     & (t,\rsu) 
\end{array} 
\end{eqnarray}
where
\begin{eqnarray}
t(T) = T \;, \qquad 
\left.\rsu(T, X)\right|^{}_{\Omega^{}_{a}} = 
\frac{\rsu_{a,\RR}-\rsu_{a,\LL}}{2}X+\frac{\rsu_{a,\LL}+ \rsu_{a,\RR}}{2}\,. \label{map2}
\end{eqnarray}
The only thing left is to prescribe the time evolution of the boundary nodes of the dynamical
domains.  When there are only two dynamical domains, as in~\cite{Canizares:2010yx}, the evolution
of their nodes is quite simple as shown above: $\rsu_{\mbox{\tiny {\rm P}-1},\LL} = const.$,  
$\rsu_{\PP,\RR} = const.$, and $\rsu_{\mbox{\tiny {\rm P}-1},\RR}=\rsu_{\PP,\LL}=\rsu_{p}(t)$.
Now, let us consider the situation where we have $N^{}_{d}$ dynamical domains at each side of
the particle, $2N^{}_{d}$ in total (Figure~\ref{multidomain} shows the case $N^{}_{d}=3$).
Following Figure~\ref{multidomain}, let us assume that the first dynamical domain is 
$\Omega^{}_{m}=[\rsu_{m,\LL},\rsu_{m,\RR}]$, which means that ${\rm P} = m + N^{}_{d}$, that is
\begin{eqnarray}
\Omega & = & [\rsu_{\HH}, \rsu_{1,\RR}]\cup\cdots\cup[\rsu_{m,\LL},\rsu_{m,\RR}]\cup\cdots
\cup[\rsu_{m+N^{}_{d}-1,\LL}, \rsu_{p}]\cup[\rsu_{p},\rsu_{\PP,\RR}]\nonumber \\ 
& \cup & \cdots\cup [\rsu_{\PP+N^{}_{d}-1,\LL}, \rsu_{\PP+N^{}_{d}-1,\RR}]\cup\cdots\cup[\rsu_{d,\LL}, \rsu_{\II}]\,.
\label{detailedgrid}
\end{eqnarray}
The criterium that we use to determine the motion of the dynamical domains is to demand that 
all of them to the left of the particle have the same coordinate size (in terms of the radial 
tortoise coordinate) at any given time, and the same for those to the right of the particle.
This completely determines the evolution of the dynamical domains.  Then, the motion of the dynamical 
nodes to the left of the particle is described by the following equations 
($a=m,\ldots,m+N^{}_{d}-1$):
\begin{eqnarray}
\rsu_{a,\LL} & = & \rsu_{m,\LL} + (a-m)\frac{\rsu_{p}-\rsu_{m,\LL}}{N^{}_{d}}\,, \\[1mm]
\rsu_{a,\RR} & = & \rsu_{m,\LL} + (a-m+1)\frac{\rsu_{p}-\rsu_{m,\LL}}{N^{}_{d}}\,,
\end{eqnarray}
and the motion of the ones to the right of the particle by ($a={\rm P},\ldots,{\rm P}+N^{}_{d}-1$)
\begin{eqnarray}
\rsu_{a,\LL} & = & \rsu_{p} + (a-{\rm P})\frac{\rsu_{\PP+N^{}_{d}-1,\RR}-\rsu_{p}}{N^{}_{d}}\,, \\[1mm]
\rsu_{a,\RR} & = & \rsu_{p} + (a-{\rm P}+1)\frac{\rsu_{\PP+N^{}_{d}-1,\RR}-\rsu_{p}}{N^{}_{d}}\,,
\end{eqnarray}
where $\rsu_{m,\LL}$ and $\rsu_{\PP+N^{}_{d}-1,\RR}$ are the fixed boundary nodes that separate
the static from the dynamical domains.  The $\rsu$-coordinate size of the dynamical domains changes
according to the motion of the particle, as illustrated in Figure~\ref{multidomain}.  

\begin{figure*}
\centering 
\includegraphics[width=0.85\textwidth]{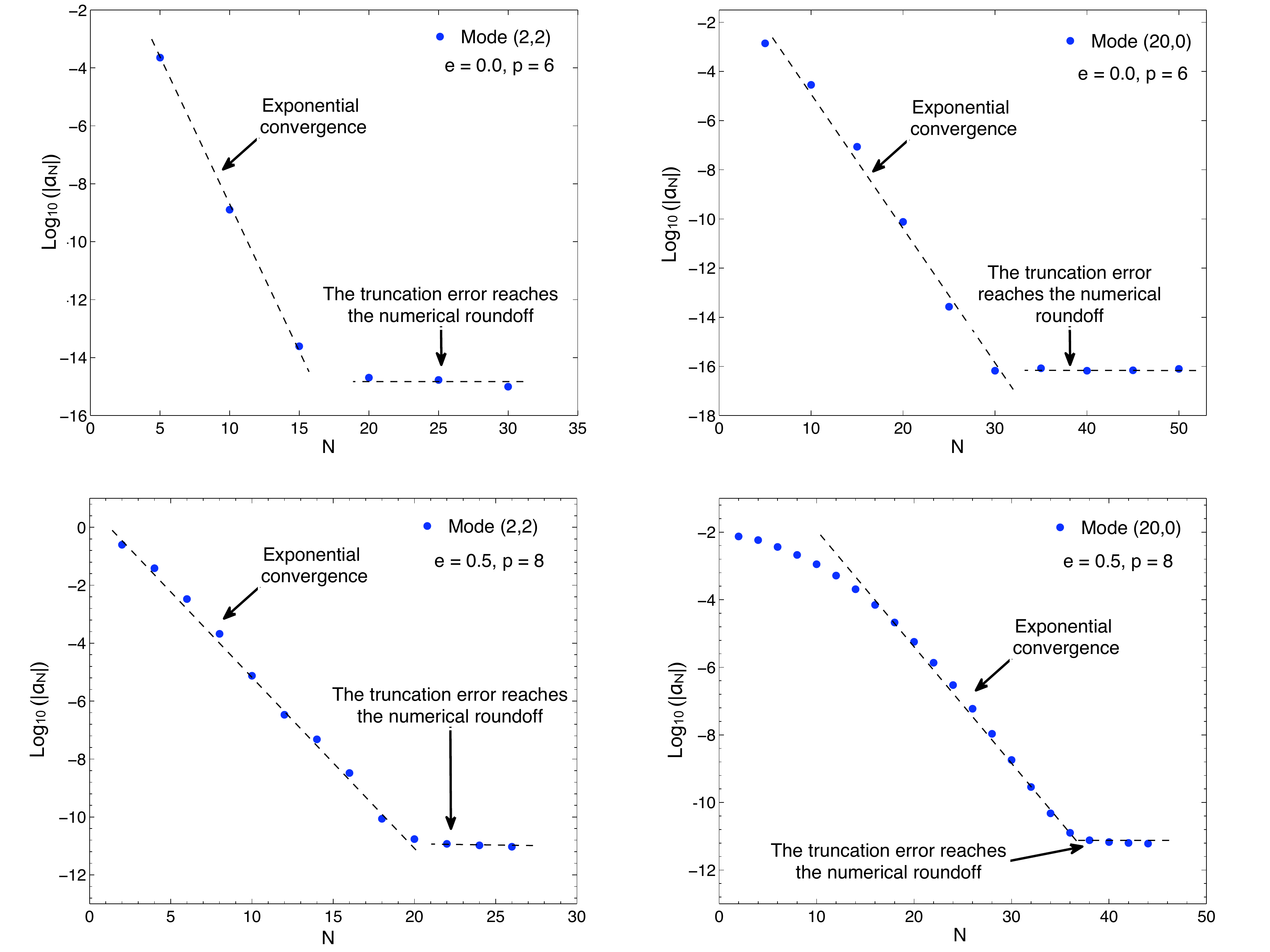}
\caption{Convergence plots ($\log_{10}|a_{N}|$ versus $N$) for the variable $\psi=r\,\Phi^{\ell m}$.
The upper row corresponds to a circular orbit with $r=6M$ (last stable orbit) and the lower row to 
an eccentric orbit with $(e,p)=(0.5,8.0)$.  The left column shows results for the harmonic mode 
$(\ell,m)=(2,2)$ and the right column for $(\ell,m) = (20,0)$. The data have been obtained from the 
domain to the right of the particle, whose coordinate size, $\Delta \rsu$, is $5M$ for the circular
case and in the eccentric case varies, due to the fact that the domain is dynamical, in the range
$1-15M$.  We can see how exponential convergence is achieved until machine roundoff is reached.
\label{convergence}}
\end{figure*}

After discretizing the domain, we need to discretize the variables and equations.  Since nothing
essential changes with respect to the description given in~\cite{Canizares:2010yx}, we just summarize
here the main points.  First we reduce the wave-type equation~(\ref{master}) to a first-order
symmetric hyperbolic system by introducing the following variables $(\psi,\psi,\varphi) = 
(r\,\Phi^{\ell m},\partial^{}_t\psi^{\ell m}\,, \partial^{}_{\rsu}\psi^{\ell m})$.  Then, the 
restriction of these variable to every single domain satisfies a set of homogeneous equations, i.e.
without Dirac delta distributions.  This means that at each domain we obtain smooth solutions.
This is the main property of the PwP scheme and of its implementation. The fact that the solutions at each
domain are smooth preserves the exponential convergence property of the PSC method.  At the same
time, the scheme avoids the presence of singularities and hence, the need of using 
implementations that introduce artificial spatial scales in the problem in order to deal with 
the particle.  We show the convergence properties of our computations in Figure~\ref{convergence}, where 
we plot an estimation of the truncation error based on the value of the last spectral coefficient,
$\mb{a}_{N}$, with respect to the number of collocation points, $N$.

The other important point of the PwP scheme is the communication of the solutions from the different 
domains across their boundaries.   This is done by using the junction conditions dictated by
our field equations (see~\cite{Sopuerta:2005gz,Canizares:2009ay,Canizares:2010yx}).  In the case
of boundaries that do not contain the particle this translates to impose continuity of the retarded
field and its first-order time and radial derivatives.  When the particle is present the junction
conditions translate into jumps in the derivatives (not in the retarded field itself).  
The junction conditions are imposed in pratice using two different methods described 
in~\cite{Canizares:2010yx}: (i) The \emph{penalty} method, and 
(ii) the direct communication of the characteristic fields.

For the spatial discretization of our variables, $\mb{U}=(\psi,\psi,\varphi)$, the PSC method provides two 
representations: (i) the {\em physical} one, based on the values of the variables at the collocation
points, $\{\mb{U}_{i}\}_{i=0,\ldots,N}$; (ii) the {\em spectral} representation, based on the 
coefficients of the truncated expansion in Chebyshev polynomials, $\{\mb{a}_n\}_{n=0,\ldots,N}$. 
We change from one representation to the other by means of transformation matrices or, in a more
efficient way, by means of fast Fourier transformations~\cite{Boyd}.  Finally, the discretization in
time, the evolution algorithm, is done using a Runge-Kutta 4 algorithm.

\section{Tuning the Domain Size to the Harmonic Number \mbox{$m$}}  \label{fitmod}

In the formulation and implementation of the PwP scheme presented in~\cite{Canizares:2009ay,Canizares:2010yx}
it was already shown that precise calculations of the self-force with modest computational resources
(less than an hour in a computer with two Quad-Core Intel Xeon processors at 2.8 GHz) is possible.   
However, as already discussed in~\cite{Canizares:2009ay,Canizares:2010yx}, there
is a room for improvement in different aspects of the numerical implementation.  Here,
we show how to improve what we believe is the most important aspect: the distribution of 
domain sizes with respect to the harmonic number $m$.  To that end, we use a phenomenological
approach to this question, restricting ourselves to the particular case of circular orbits.  

The scalar field harmonic modes produced by a charged particle in circular motion are 
like {\em forced} oscillators, with a force term given by $S^{\ell m}\,\delta (r-r^{}_{p})$ 
[see Eq.~(\ref{master})], which oscillates periodically in time.  More precisely,
$S^{\ell m}\propto\exp(i\,m\,\omega_p\,t)$, where $\omega_p = M^{1/2} r_p^{-3/2}$ is the particle's 
angular velocity.  The potential $V^{}_{\ell}$ decays quite fast away from the location of its
maximum, so that the field looks like a monocromatic wave in the regions where the potential is not 
important.  The wavelength of these waves is $\lambda_m = \lambda_{1}/m$, where 
$\lambda_{1}= 2\pi/\omega_p$ is the wavelength of the $m = 1$ modes.  Therefore in a domain
$\Omega_{a}$, with coordinate size $\Delta\rsu = \rsu_{a,\RR} - \rsu_{a,\LL}$, modes with different
harmonic number $m$ require different resolutions, understood as the ratio $\Delta\rsu/N$, in
order to resolved them with the same accuracy. For instance, given two modes $\Phi^{\ell m}$ and
$\Phi^{\ell' m'}$ such that $m < m'$, i.e. with $\lambda_m > \lambda_m'$, in a given domain 
$\Omega_{a}$ we will find more wavelengths of the mode $\Phi^{\ell' m'}$ than the mode 
$\Phi^{\ell m}$.  This means we need more resolution for the mode  $\Phi^{\ell' m'}$ than for
the mode  $\Phi^{\ell m}$ in order to achieve comparable accuracy.  

\begin{figure}[htp]
\centering 
\includegraphics[width=0.45\textwidth]{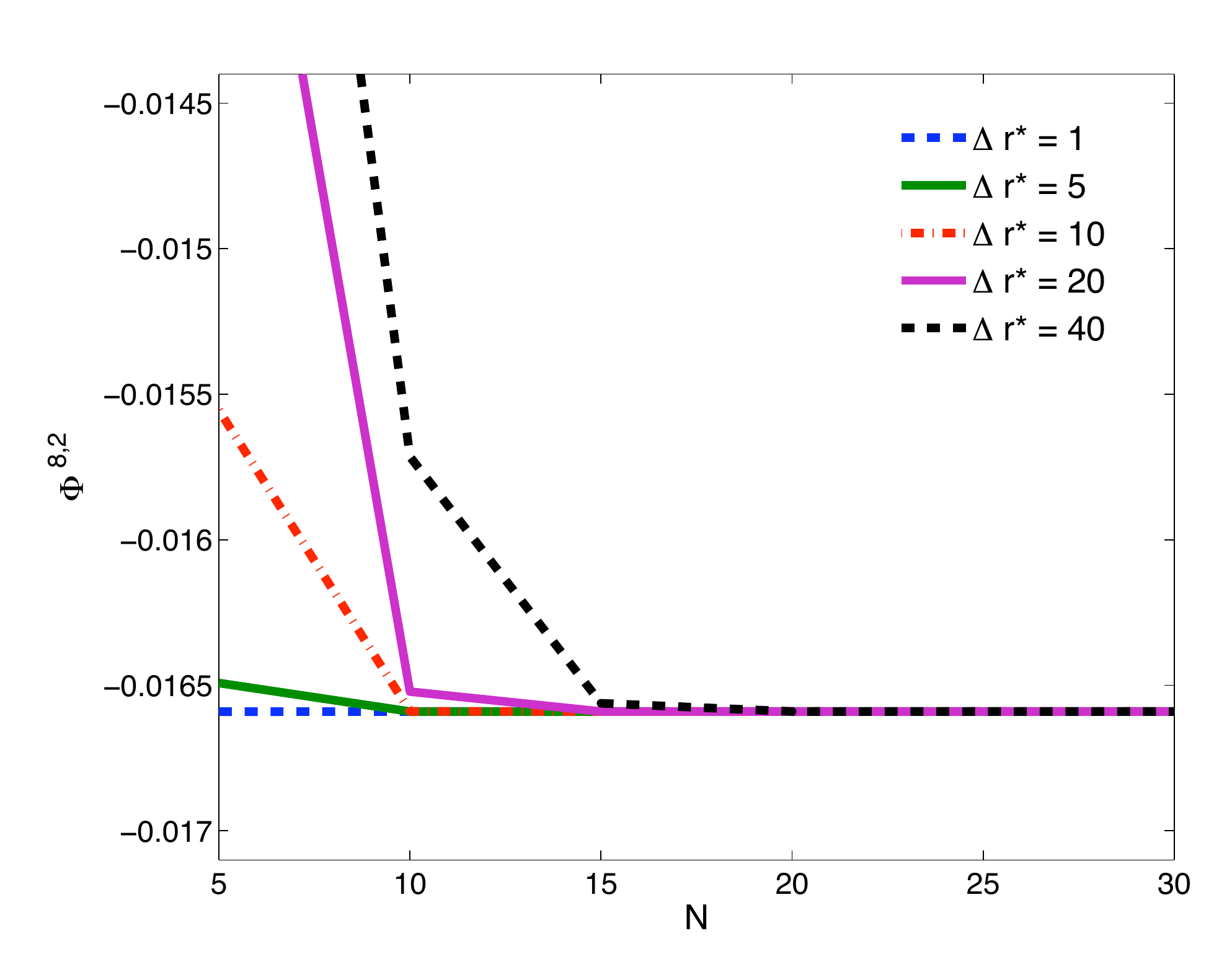}
\includegraphics[width=0.45\textwidth]{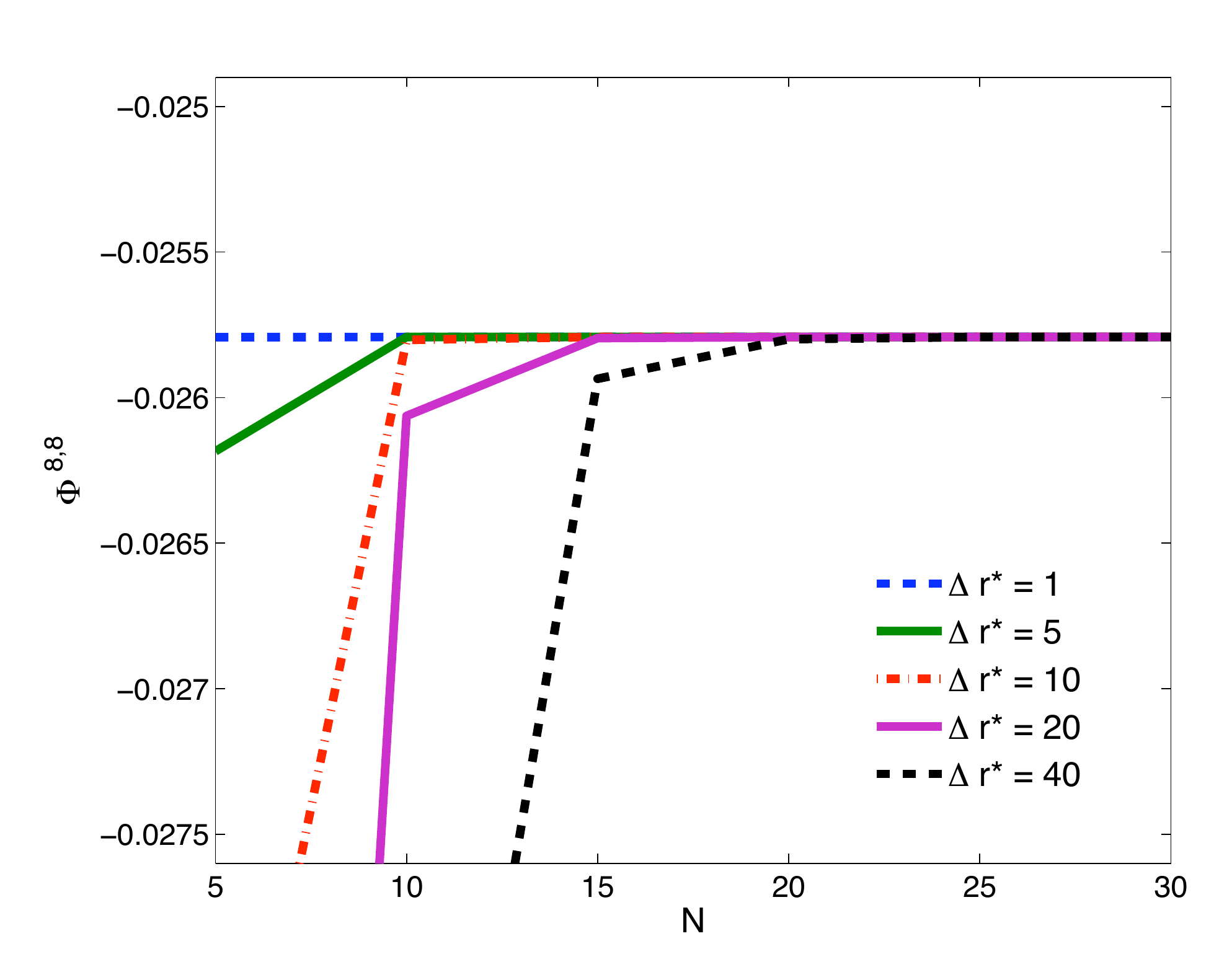}
\caption{This figure shows values of the $\Phi^{8,2}$ (left) and $\Phi^{8,8}$ (right) modes of 
the retarded field at the particle location as computed from the domain to the right of it 
($\Omega_{\PP}$), with respect to the number of collocation points $N$.  Each line correspond 
to a different coordinate size of the domain where the calculations are done: 
$\Delta \rsu/M = 1, 5, 10, 20, 40$. 
It shows how the values of the modes converge as $\Delta\rsu$ is decreased and $N$ is increased.
It is remarkable to realize that for small coordinate size the convergence is reached with few
collocation points.\label{mode_study}}
\end{figure}

Here, we can change $\Delta\rsu$ and the number of domains $d$ (if we reduce $\Delta\rsu$ we
need more domains and the converse), or $N$, or both.  
The question here is what is the optimum way of setting the resolution for each $m$-mode,
in the sense of minimizing the computational resources.  
Increasing the resolutions means to increase the number of computer operations, and in this sense
it is not the same to increase the number of domains, decreasing their size (see~\cite{Canizares:2009ay} 
for a discussion of this point), than to increase the number of collocation points (using FFT 
instead of matrices, the number of operations in the relevant part of the algorithm scales 
with $N$ as $N\ln N$).   In order to investigate this question we have conducted a series of simulations
to study the behavior of different modes of the retarded field $\Phi^{\ell m}$  
under changes of the coordinate domain size $\Delta\rsu$ and the number of collocation points $N$.
The results for values at the particle location for the
modes $\Phi^{8,2}$ and $\Phi^{8,8}$ are shown in Figure~\ref{mode_study}. 
As we can see, it is crucial to choose carefully $\Delta \rsu$ for a fixed value of $N$, otherwise
we can find ourselves in one of the following situations: Either $\Delta \rsu$ is too big and then, 
$N$ collocation points are not enough to resolve all the modes, or $\Delta \rsu$ is too small and  
we are using too many collocation points, i.e. we are performing too many numerical computations
to resolve the modes. 

From our simulations we have found that adjusting the size of the domains to the wavelength of the modes
of the retarded field, i.e. $\Delta\rsu \simeq \lambda_m$, the values of the field converge and
are resolved for a minimum of collocation points given by $N \sim \Delta \rsu/(2M) + 25$.  
This provides us with a rule of thumb for the choice of the domain size and number of collocation
points as a function of $m$:  $\Delta\rsu(m) \sim \lambda_m$  and  
$N(m) = [\Delta\rsu/(2M) +25]$ (where here $[x]$ denotes the integer closest to $x$). 

Another strategy that has been used in the literature (see~\cite{DiazRivera:2004ik,Thornburg:2010tq})
is to truncate the mode series for the self-force at a given $\ell = \ell_{\TT}$ and then, to estimate
the remaining infinite number of terms by using the known large-$\ell$ series~\cite{Detweiler:2002gi}.  
The idea is to fit the coefficients of this series by least-squares fitting to a subset of the 
numerically estimated self-force modes (i.e. with $\ell \leq \ell_{\TT}$). We have not used this 
procedure in this paper because our aim is to tune our numerical scheme, but it can certainly be used 
in order to complement and improve the results that we obtain.

\section{Conclusions and Discussion}\label{discussion}

The PwP scheme introduced and developed in~\cite{Canizares:2008dp,Canizares:2009ay,Canizares:2010yx} 
provides a framework for precise and efficient computations of the self-force in the time
domain.  In this paper we have investigated how to tune the method in order to increase its efficiency 
and accuracy.  We have also extended the PwP scheme in order to make it suitable for computations
of the self-force on highly eccentric orbits.

We have studied how to distribute the domain sizes and the number of collocation points
so that we allocate the optimum resolution to each harmonic mode.  This is
very important as the resolution requirement increases with $m$, 
despite the fact that  high $\ell$-modes contribute less to the self-force
than low $\ell$ ones.  This means that in order to improve the accuracy of the self-force we must 
have a good control of the resolution, otherwise modes with high $m$ (and hence with high $\ell$)
will limit the precision despite not contributing much to the self-force.  

Using this information, we have conducted a series of computations of the self-force
for the case of a scalar charged particle in circular (geodesic) orbits around a non-rotating
BH.   The computational parameter space that we have explored is described as follows:  
The total number of domains that we have used is in the range $d=20-43$.  The coordinate size of the 
large domains (the ones far from the particle and near the boundaries) is in the range
$\Delta\rsu =50 - 100M$.  The number of collocation points has been fixed to $N = 50$.  
The largest $\ell$, $\ell_{\MAX}$, considered in the computations is in the range $\ell_{\MAX}=20-40$.

Adapting the resolution for each $m$-mode also allows us to adapt the time step as the CFL
condition is proportional to the minimum coordinate physical distance (as measured in terms
of the tortoise radial coordinate) between grid points (and inversely
proportional to the square of the number of collocation points).  This means that the
number of time steps required, $N_{t}$, for a simulation is going to be proportional to $m$ as:
$N_{t}^m = m\,N_{t}^{m=1}$.    This is assuming that we use the same number of domains for all
modes, but it is clear that modes with low $m$ would need less domains than modes with
high $m$, and therefore, this is another source of reduction of computational time.

We have been able to obtain very  precise values of the self-force for a wide range of values 
within the parameter space.  For instance, in the case of the radial component of the regular field, $\Phi^{\RR}_r$,
the only one that for the circular case needs regularization, we have obtained
values like $\Phi^{\RR}_r = 1.677282\times10^{-4}\, q/M^2\,,$ which
have a relative error of the order of $5\cdot10^{-5}\,\%$ with respect to the values obtained 
in~\cite{DiazRivera:2004ik} using frequency-domain methods.  The relative error is always
in the range $5\cdot10^{-5}\,\% - 5\cdot10^{-3}\,\%$.  This constitutes a significant 
improvement with respect to our own previous estimations presented in~\cite{Canizares:2009ay},
where the relative errors quoted for $\Phi^{\RR}_r$ were of the order of $0.1\%$.  
The typical time for a full self-force calculation in a computer with two Quad-Core Intel Xeon 
processors at $2.27$ GHz is in the range $10-15$ minutes, which is a very significant reduction with 
respect to our computations in~\cite{Canizares:2009ay}, specially taking into account that we 
have also  improved the precision of the self-force computations.

The calculations we have presented can be further improved in terms of computational
time, and perhaps in accuracy by exploring techniques to bring the boundaries closer
to the particle without degrading the accuracy of the field values near it.
This can be done either by improving the outgoing boundary conditions (see, e.g.~\cite{Lau:2004as}) or by using
some sort of compactification of the physical domain (see, e.g.~\cite{2010arXiv1008.3809Z}).  
Another possibility for making the computations faster (although this does not decrease the CPU time) 
is to parallelize the code and use computers with many cores.  This is in principle a simple task as the
different modes are not coupled.  In any case, the next step in this line of work is to perform 
a similar phenomenological study for the eccentric case, where not only the azimuthal frequency is important, 
but also the radial one, which is absent in the circular case.

\section*{Acknowledgments}
	
P.C.M. is supported by a predoctoral FPU fellowship of the Spanish Ministry of
Science and Innovation (MICINN).
C.F.S. acknowledges support from the Ram\'on y Cajal Programme of the
Ministry of Education and Science (MEC) of Spain, by a Marie Curie
International Reintegration Grant (MIRG-CT-2007-205005/PHY) within the
7th European Community Framework Programme (EU-FP7), and contracts
ESP2007-61712 (MEC) and No.~FIS2008-06078-C03-01/FIS (MICINN). We acknowledge
the computational resources provided by the Centre de Supercomputaci\'o de
Catalunya (CESCA) and the Centro de Supercomputaci{\'o}n de Galicia 
(CESGA: ICTS-2009-40).

\section*{References}

\providecommand{\newblock}{}

\end{document}